\documentclass[a4paper,english,aps,manuscript]{revtex4}
\usepackage[T1]{fontenc}
\usepackage[latin9]{inputenc}
\usepackage{amsmath}

\makeatletter

\newcommand{\noun}[1]{\textsc{#1}}

\usepackage{amsfonts}

\usepackage{bbm}

\usepackage{amscd}

\usepackage{array}

\usepackage{tensind}

\usepackage{mathrsfs}

\tensordelimiter{?}

\usepackage{longtable}

\def\laq{\raise 0.4ex\hbox{$<$}\kern -0.8em\lower 0.62ex\hbox{$\sim$}}
\def\gaq{\raise 0.4ex\hbox{$>$}\kern -0.7em\lower 0.62ex\hbox{$\sim$}}

\newcommand{\ba}{\begin{array}}
\newcommand{\ea}{\end{array}}

\newcommand{\be}{\begin{equation}}
\newcommand{\ee}{\end{equation}}
\newcommand{\bea}{\begin{eqnarray}}
\newcommand{\eea}{\end{eqnarray}}

\newcommand{\mytextrm}[1]{{}}

\def\ee{{\mathrm{e}}}

\newlength{\sizeonefig}
\newlength{\sizetwofig}
\newlength{\sizeonefigb}
\newlength{\sizetwofigb}
\makeatother

\usepackage{babel}

\begin{document}

\title{Binary spinning black hole Hamiltonian in canonical center-of-mass and rest-frame
coordinates through higher post-Newtonian order}

\author{Tilman J. Rothe}

\affiliation{ Mathematisches Institut, Friedrich-Schiller-Universit\"at Jena,
Ernst-Abbe-Platz 2, 07743 Jena, Germany}

\author{Gerhard Sch\"afer}

\affiliation{Theoretisch-Physikalisches Institut, Friedrich-Schiller-Universit\"at Jena, 
Max-Wien-Platz 1, 07743 Jena, Germany}

\begin{abstract}
The recently constructed Hamiltonians for spinless binary black holes through
third post-Newtonian order and for spinning ones through
formal second post-Newtonian order, where the spins are counted of
zero post-Newtonian order, are transformed into fully canonical
center-of-mass and rest-frame variables.
The mixture terms in the Hamiltonians between center-of-mass and rest-frame variables
are in accordance with the relation between the total linear momentum
and the center-of-mass velocity as demanded by global Lorentz invariance.
The various generating functions for the center-of-mass and rest-frame canonical
variables are explicitly given in terms of the single-particle canonical variables. The no-interaction 
theorem does not apply because the world-line condition of Lorentz covariant 
position variables is not imposed.

\vspace{2ex}\noindent
PACS number(s): 11.30.Cp, 04.25.Nx, 45.20.Jj, 45.50.Pk
\end{abstract}


\date{\today}

\maketitle
\newcommand{\clight}{c}
\newcommand{\assign}{\equiv}
\newcommand{\op}[1]{\operatorname{#1}}
\newcommand{\fett}[1]{\boldsymbol{\bold{#1}}}
\newcommand{\kreu}{\fett{\times}}
\newcommand{\parenmacro}[1]{\left(#1\right)}
\newcommand{\parenMacroInvisible}[1]{ #1}
\newcommand{\PB}[2]{\left\{  #1,\,#2\right\}  }
\newcommand{\noop}[1]{#1}
\newcommand{\pqm}{1}
\newcommand{\pqe}{2}
\newcommand{\pqh}{3}
\newcommand{\pqr}{R}
\newcommand{\pqs}{r}
\newcommand{\pqme}{c}
\newcommand{\pqmeh}{C}
\newcommand{\mc}{m_{\pqm}+m_{\pqe}+m_{\pqh}}
\newcommand{\mme}{m_{\pqm}+m_{\pqe}}
\newcommand{\mh}{m_{\pqh}}
\newcommand{\mm}{m_{\pqm}}
\newcommand{\me}{m_{\pqe}}
\newcommand{\muc}{\mu_{\pqmeh}}
\newcommand{\mur}{\mu_{\pqr}}
\newcommand{\mus}{\mu}
\newcommand{\muz}{m}
\newcommand{\lpar}{\bigg(}
\newcommand{\lbra}{\bigg[}
\newcommand{\lcur}{\bigg\{}
\newcommand{\rpar}{\bigg)}
\newcommand{\rbra}{\bigg]}
\newcommand{\rcur}{\bigg\}}
\newcommand{\myvec}[1]{\fett{#1}}
\newcommand{\boost}{\myvec{G}}
\newcommand{\pcabs}{p_{\pqmeh}}
\newcommand{\pc}{\myvec{p}_{\pqmeh}}
\newcommand{\prabs}{p_{\pqr}}
\newcommand{\pr}{\myvec{p}_{\pqr}}
\newcommand{\psabs}{p}
\newcommand{\ps}{\myvec{p}}
\newcommand{\xcabs}{x_{\pqmeh}}
\newcommand{\xc}{\myvec{x}_{\pqmeh}}
\newcommand{\xrabs}{x_{\pqr}}
\newcommand{\xr}{\myvec{x}_{\pqr}}
\newcommand{\rr}{r_{\pqr}}
\newcommand{\xsabs}{x}
\newcommand{\xs}{\myvec{x}}
\newcommand{\rs}{r}
\newcommand{\vpm}{\myvec{p}_{\pqm}}
\newcommand{\pmabs}{p_{\pqm}}
\newcommand{\vpe}{\myvec{p}_{\pqe}}
\newcommand{\peabs}{p_{\pqe}}
\newcommand{\vph}{\myvec{p}_{\pqh}}
\newcommand{\phabs}{p_{\pqh}}
\newcommand{\vxm}{\myvec{x}_{\pqm}}
\newcommand{\xmabs}{x_{\pqm}}
\newcommand{\vxe}{\myvec{x}_{\pqe}}
\newcommand{\xeabs}{x_{\pqe}}
\newcommand{\xz}{\myvec{X}}
\newcommand{\xzabs}{X}
\newcommand{\pzabs}{P}
\newcommand{\pz}{\myvec{P}}
\newcommand{\vxh}{\myvec{x}_{\pqh}}
\newcommand{\xhabs}{x_{\pqh}}
\newcommand{\xme}{\xz}
\newcommand{\vns}{\myvec{n}}
\newcommand{\rme}{r_{\pqm\pqe}}
\newcommand{\ns}{\vns}
\newcommand{\rmh}{r_{\pqh\pqm}}
\newcommand{\rhm}{\rmh}
\newcommand{\reh}{r_{\pqe\pqh}}
\newcommand{\rz}{R}
\newcommand{\nz}{\myvec{N}}
\newcommand{\nme}{\myvec{n}_{\pqm\pqe}}
\newcommand{\nmepart}{\nme}
\newcommand{\distm}{\rho_{\pqm}}
\newcommand{\diste}{\rho_{\pqe}}
\newcommand{\nr}{\myvec{n}_{\pqr}}
\newcommand{\vnme}{\nmepart}
\newcommand{\sm}{\myvec{S}_{\pqm}}
\newcommand{\smabs}{S_{\pqm}}
\newcommand{\se}{\myvec{S}_{\pqe}}
\newcommand{\seabs}{S_{\pqe}}
\newcommand{\neh}{\myvec{n}_{\pqe\pqh}}
\newcommand{\xeh}{\myvec{x}_{\pqe\pqh}}
\newcommand{\xhm}{\myvec{x}_{\pqh\pqm}}
\newcommand{\xmepart}{\myvec{x}_{\pqm\pqe}}
\newcommand{\nhm}{\myvec{n}_{\pqh\pqm}}
\newcommand{\psprimeabs}{\hat{p}}
\newcommand{\psprime}{\myvec{\hat{p}}}
\newcommand{\nsprime}{\myvec{\hat{n}}}
\newcommand{\rsprime}{\hat{r}}
\newcommand{\xsprime}{\myvec{\hat{x}}}
\newcommand{\scmprime}{\myvec{\hat{\mathcal{S}}}_{\pqm}}
\newcommand{\scmabsprime}{\hat{\mathcal{S}}_{\pqm}}
\newcommand{\sceprime}{\myvec{\hat{\mathcal{S}}}_{\pqe}}
\newcommand{\sceabsprime}{\hat{\mathcal{S}}_{\pqe}}
\newcommand{\scm}{\myvec{\mathcal{S}}_{\pqm}}
\newcommand{\scmabs}{\mathcal{S}_{\pqm}}
\newcommand{\sce}{\myvec{\mathcal{S}}_{\pqe}}
\newcommand{\sceabs}{\mathcal{S}_{\pqe}}
\newcommand{\pcpc}{\parenMacroInvisible{\pcabs^{2}}}
\newcommand{\pcpr}{\parenmacro{\pc\cdot\pr}}
\newcommand{\prpr}{\parenMacroInvisible{\prabs^{2}}}
\newcommand{\pcps}{\parenmacro{\pc\cdot\ps}}
\newcommand{\prps}{\parenmacro{\pr\cdot\ps}}
\newcommand{\psps}{\parenMacroInvisible{\psabs^{2}}}
\newcommand{\pmpm}{\parenMacroInvisible{\pmabs^{2}}}
\newcommand{\pepe}{\parenMacroInvisible{\peabs^{2}}}
\newcommand{\pmpe}{\parenmacro{\vpm\cdot\vpe}}
\newcommand{\pzpz}{\parenMacroInvisible{\pzabs^{2}}}
\newcommand{\pzps}{\parenmacro{\pz\cdot\ps}}
\newcommand{\xspz}{\parenmacro{\xs\cdot\pz}}
\newcommand{\nspz}{\parenmacro{\vns\cdot\pz}}
\newcommand{\nsps}{\parenmacro{\vns\cdot\ps}}
\newcommand{\xmpm}{\parenmacro{\vxm\cdot\vpm}}
\newcommand{\xmpe}{\parenmacro{\vxm\cdot\vpe}}
\newcommand{\xepe}{\parenmacro{\vxe\cdot\vpe}}
\newcommand{\xepm}{\parenmacro{\vxe\cdot\vpm}}
\newcommand{\pexm}{\parenmacro{\vxm\cdot\vpe}}
\newcommand{\pmxm}{\parenmacro{\vxm\cdot\vpm}}
\newcommand{\pmxe}{\parenmacro{\vxe\cdot\vpm}}
\newcommand{\pexe}{\parenmacro{\vxe\cdot\vpe}}
\newcommand{\xsps}{\parenmacro{\xs\cdot\ps}}
\newcommand{\psxz}{\parenmacro{\ps\cdot\xz}}
\newcommand{\nsxz}{\parenmacro{\ns\cdot\xz}}
\newcommand{\xzxz}{\parenmacro{\xzabs^{2}}}
\newcommand{\pzxz}{\parenmacro{\pz\cdot\xz}}
\newcommand{\nzpz}{\parenmacro{\nz\cdot\pz}}
\newcommand{\nzns}{\parenmacro{\nz\cdot\ns}}
\newcommand{\nzps}{\parenmacro{\nz\cdot\ps}}
\newcommand{\phph}{\parenmacro{\phabs^{2}}}
\newcommand{\pmnme}{\parenmacro{\vpm\cdot\nmepart}}
\newcommand{\penme}{\parenmacro{\vpe\cdot\nmepart}}
\newcommand{\nmepepart}{\parenmacro{\nmepart\cdot\vpe}}
\newcommand{\nmepm}{\parenmacro{\nmepart\cdot\vpm}}
\newcommand{\nmepe}{\parenmacro{\nmepart\cdot\vpe}}
\newcommand{\nrps}{\parenmacro{\nr\cdot\ps}}
\newcommand{\smnme}{\parenmacro{\sm\cdot\nmepart}}
\newcommand{\smsm}{\parenMacroInvisible{\smabs^{2}}}
\newcommand{\senme}{\parenmacro{\se\cdot\nmepart}}
\newcommand{\smse}{\parenmacro{\sm\cdot\se}}
\newcommand{\epsnspsxz}{\parenmacro{\ns\times\ps\cdot\xz}}
\newcommand{\nehph}{\parenmacro{\neh\cdot\vph}}
\newcommand{\nmeph}{\parenmacro{\nmepart\cdot\vph}}
\newcommand{\nehpm}{\parenmacro{\neh\cdot\vpm}}
\newcommand{\xehph}{\parenmacro{\xeh\cdot\vph}}
\newcommand{\xehpm}{\parenmacro{\xeh\cdot\vpm}}
\newcommand{\xmeph}{\parenmacro{\xmepart\cdot\vph}}
\newcommand{\xmepm}{\parenmacro{\xmepart\cdot\vpm}}
\newcommand{\pmph}{\parenmacro{\vpm\cdot\vph}}
\newcommand{\xehpe}{\parenmacro{\xeh\cdot\vpm}}
\newcommand{\xmepe}{\parenmacro{\xmepart\cdot\vpe}}
\newcommand{\nrpr}{\parenmacro{\nr\cdot\pr}}
\newcommand{\cosangle}{\cphi}
\newcommand{\nspr}{\parenmacro{\ns\cdot\pr}}
\newcommand{\nspc}{\parenmacro{\ns\cdot\pc}}
\newcommand{\nssm}{\parenmacro{\ns\cdot\sm}}
\newcommand{\nsse}{\parenmacro{\ns\cdot\se}}
\newcommand{\epsnspssm}{\parenmacro{\ns\cdot\ps\kreu\sm}}
\newcommand{\epsnspsse}{\parenmacro{\ns\cdot\ps\kreu\se}}
\newcommand{\sese}{\parenmacro{\seabs^{2}}}
\newcommand{\cphi}{\parenmacro{\ns\cdot\nr}}
\newcommand{\pspsprime}{\parenMacroInvisible{\psprimeabs^{2}}}
\newcommand{\nspsprime}{\parenmacro{\nsprime\cdot\psprime}}
\newcommand{\epspmpeSm}{\parenmacro{\vpm\cdot\vpe\kreu\sm}}
\newcommand{\epspmpeSe}{\parenmacro{\vpm\cdot\vpe\kreu\se}}
\newcommand{\nmeSm}{\parenmacro{\nme\cdot\sm}}
\newcommand{\nmeSe}{\parenmacro{\nme\cdot\se}}
\newcommand{\pmSm}{\parenmacro{\vpm\cdot\sm}}
\newcommand{\pmSe}{\parenmacro{\vpm\cdot\se}}
\newcommand{\SmSm}{\parenMacroInvisible{\smabs^{2}}}
\newcommand{\SmSe}{\parenmacro{\sm\cdot\se}}
\newcommand{\SeSe}{\parenMacroInvisible{\seabs^{2}}}
\newcommand{\epsnmepmSm}{\parenmacro{\nme\cdot\vpm\kreu\sm}}
\newcommand{\epsnmepmSe}{\parenmacro{\nme\cdot\vpm\kreu\se}}
\newcommand{\epsnmepeSm}{\parenmacro{\nme\cdot\vpe\kreu\sm}}
\newcommand{\peSm}{\parenmacro{\vpe\cdot\sm}}
\newcommand{\peSe}{\parenmacro{\vpe\cdot\se}}
\newcommand{\epsnmepeSe}{\parenmacro{\nme\cdot\vpe\kreu\se}}
\newcommand{\nssceprime}{\parenmacro{\nsprime\cdot\sceprime}}
\newcommand{\nsscmprime}{\parenmacro{\nsprime\cdot\scmprime}}
\newcommand{\pssceprime}{\parenmacro{\psprime\cdot\sceprime}}
\newcommand{\epsnspssceprime}{\parenmacro{\nsprime\cdot\psprime\kreu\sceprime}}
\newcommand{\psscmprime}{\parenmacro{\psprime\cdot\scmprime}}
\newcommand{\epsnspsscmprime}{\parenmacro{\nsprime\cdot\psprime\kreu\scmprime}}
\newcommand{\scmsceprime}{\parenmacro{\scmprime\cdot\sceprime}}
\newcommand{\epsnsscmsceprime}{\parenmacro{\nsprime\cdot\scmprime\kreu\sceprime}}
\newcommand{\epspsscmsceprime}{\parenmacro{\psprime\cdot\scmprime\kreu\sceprime}}
\newcommand{\scmscmprime}{\parenMacroInvisible{\scmabsprime^{2}}}
\newcommand{\scesceprime}{\parenMacroInvisible{\sceabsprime^{2}}}
\newcommand{\nsscm}{\parenmacro{\ns\cdot\scm}}
\newcommand{\nssce}{\parenmacro{\ns\cdot\sce}}
\newcommand{\scmscm}{\parenMacroInvisible{\scmabs^{2}}}
\newcommand{\scmsce}{\parenmacro{\scm\cdot\sce}}
\newcommand{\epsnspsscm}{\parenmacro{\ns\cdot\ps\kreu\scm}}
\newcommand{\psscm}{\parenmacro{\ps\cdot\scm}}
\newcommand{\pssce}{\parenmacro{\ps\cdot\sce}}
\newcommand{\scesce}{\parenMacroInvisible{\sceabs^{2}}}
\newcommand{\epsnspssce}{\parenmacro{\ns\cdot\ps\kreu\sce}}
\newcommand{\epsnsscmsce}{\parenmacro{\ns\cdot\scm\kreu\sce}}
\newcommand{\epspsscmsce}{\parenmacro{\ps\cdot\scm\kreu\sce}}
\newcommand{\epsnspzscm}{\parenmacro{\ns\cdot\pz\kreu\scm}}
\newcommand{\epspzpsscm}{\parenmacro{\pz\cdot\ps\kreu\scm}}
\newcommand{\pzscm}{\parenmacro{\pz\cdot\scm}}

\section{Introduction}

The conservative Hamiltonians for non-spinning and spinning
binary black holes are known to higher post-Newtonian (PN) orders in
global inertial reference frames
\citep{JS98,DJS00,DJS00aErr,DJS01dimReg,DJS08a,SteiSchaHerg08,HergScha08a,HergScha08b,
SteiHergScha08}. However, applications are typically made in the
rest frame where the total linear momentum of the binary system 
vanishes \citep{DJS00c,DJS00b,DJS08b}.
In that case six degrees of freedom in phase space get suppressed
and the remaining Hamiltonians simplify a lot. By boosting the system
back, the original Hamiltonians are recovered but within a set of
canonical variables different from the former single-particle ones. 
If a Hamiltonian were given in Lorentz covariant
coordinates, boosting would be straightforward. However, the
canonical coordinates the Hamiltonians are presented in are quite
different from Lorentz covariant ones, so boosting of those coordinates is quite an involved
procedure. 
In this paper we will solve the problem of boosted representation
of Hamiltonians by constructing fully canonical
center-of-mass and rest-frame coordinates. This work is based on the
diploma thesis of one of the authors \citep{Rot09}. 

A historical view to the relativistic mechanics in Hamiltonian form, see,
e.g. \citep{AlbCraLus07}, clearly shows that the problem of
constructing center-of-mass and rest-frame coordinates for interacting
relativistic systems has a long tradition beginning with the pioneering works by
Thomas, Bakamjian, and Foldy around the 1950s of the last century \citep{Tho52,BakTho53,Fol61}.
To our best knowledge, the known general relativistic gravitational Hamiltonians
for binary point-like particles with and without spin have never been given
in canonical center-of-mass and rest-frame coordinates, not even at
the 1PN level where the world-line condition still applies, i.e. where fully canonical coordinates and Lorentz 
covariance of particle position vectors are still compatible, see,
e.g. \citep{StaHav76, PauPro76}.  Evidently, the center-of-mass and
relative position coordinates we are searching for are of Newton-Wigner type because of component-wise vanishing
Poisson brackets \citep{NewWig49}.

The importance of the present work can be seen in two directions (i),
it is by far not trivial to explicitly construct the involved canonical coordinate
transformations in phase space and (ii), for future applications, the ground will be led 
for a  Hamiltonian treatment of e.g. the recoil in binary systems
through gravitational radiation emission or of the orbital motion or scattering of binaries in
many-body systems, see, e.g., \cite{Brum91} for the post-Newtonian motion of the Moon or 
\citep{Vino72} for a relativistic generalization of the Jacobi momenta of the non-relativistic three-body problem.

\section{Implications from the Poincar\'e algebra}

In space-asymptotically flat spacetimes, the ten conserved quantities
total energy $H$, linear momentum $P^{i}\equiv P_{i}$,
angular momentum $J^{i}\equiv J_{i}$, and Lorentz boost vector
$K^{i}$ have to fulfill the Poincar\'e algebra
\begin{subequations}\label{eq:poincare algebra klammern}\begin{equation}
\PB{P^{i}}{H}=\PB{J^{i}}{H}=0,\end{equation}
 \begin{equation}
\PB{G^{i}}{H}=P^{i},\quad\PB{P^{i}}{P^{j}}=0,\end{equation}
\begin{equation}
\PB{J^{i}}{P^{j}}=\epsilon^{ijk}P^{k},\quad\PB{J^{i}}{J^{j}}=\epsilon^{ijk}J^{k},\end{equation}
\begin{equation}
\PB{J^{i}}{G^{j}}=\epsilon^{ijk}G^{k},\quad\PB{G^{i}}{P^{j}}=\frac{1}{\clight^{2}}H\delta^{ij},\end{equation}
\begin{equation}
\PB{G^{i}}{G^{j}}=-\frac{1}{\clight^{2}}\epsilon^{ijk}J^{k},\end{equation}\end{subequations}
where $G^{i}=K^{i} + P^{i}t$ is the non-explicitly time-dependent center-of-energy vector $G^{i}$.

Defining \begin{equation}
M=\frac{1}{\clight^{2}}\sqrt{H^{2}-\clight^{2}\pzpz},\quad\mbox{or},\quad H=\sqrt{\clight^{4}M^{2}+\clight^{2}\pzpz},\end{equation}
a canonical center-of-mass coordinate $X^{i}$ can be introduced
in the following manner \citep{Gal65,Hill67,LorRom74,SW09} \begin{subequations}\begin{eqnarray}
\myvec{X} & = & \clight^{2}\frac{{\bf
\myvec{G}}}{H}+\frac{1}{M\left(H+c^{2}M\right)}
\left(\myvec{J}-c^{2}\left(\frac{{\bf \myvec{G}}}{H}\kreu{\bf
\pz}\right)\right)\kreu\pz,\label{eq:Newton-Wigner-Schwerpunkt}
\end{eqnarray}
 where $\myvec{G}=(G^{i})$ etc. Reversely, the center-of-energy vector
can be expressed in the form \begin{equation}
\myvec{G}=\frac{1}{\clight^{2}}H\xz-\frac{1}{M\clight^{2}+H}\left(\myvec{J}-\xz\kreu\pz\right)\kreu\pz.\end{equation}\end{subequations}
It is crucial to point out that we make no further assumptions on $\xz$ or $\myvec{G}$.

The fundamental Poisson brackets read \begin{subequations}\begin{eqnarray}
\PB{X^{i}}{X^{j}} & = & 0,\\
\PB{X^{i}}{P^{j}} & = & \delta^{ij},\\
\PB{P^{i}}{P^{j}} & = & 0.\end{eqnarray}\end{subequations}
 Further, for the center-of-mass velocity ${\bf V}$, \begin{eqnarray}
{\bf \myvec{V}}\equiv\dot{{\bf \xz}}=\PB{\myvec{X}}{H}=\frac{\clight^{2}\pz}{H}, & \qquad\mbox{or},\qquad & {\bf \pz}=\frac{H}{\clight^{2}}\myvec{V}\end{eqnarray}
 are valid. Hereof the relations  \begin{eqnarray}
\pz  =  \frac{M\myvec{V}}{\sqrt{1-\frac{V^{2}}{\clight^{2}}}} \quad  \mbox{and} \quad
H  =  \frac{Mc^2}{\sqrt{1-\frac{V^{2}}{\clight^{2}}}} \end{eqnarray}
follow as the single-particle case mandates. Notice that it also holds
\begin{eqnarray}
{\bf \myvec{V}} = \PB{\myvec{G}}{H}\frac{c^2}{H},\end{eqnarray} i.e. the speed of the center-of-energy vector
coincides with the speed of the canonical center-of-mass position vector.

The Poisson brackets resulting from the Poincar\'e algebra, \begin{subequations}\begin{eqnarray}
\PB{H}{M} & = & 0,\\
\PB{P^{i}}{M} & = & 0,\\
\PB{X^{i}}{M} & = & 0, \end{eqnarray}\end{subequations}
are crucial in the following. They fully generally show that the rest-mass energy $M$
is a constant of motion not depending on the center-of-mass phase-space
coordinates ${\bf X}$ and ${\bf P}$.

Restricting ourselves at the beginning to spinless binary point-mass systems, the
explicit expressions for the total linear and orbital angular momentum read \begin{eqnarray}
{\bf \pz}=\vpm+\vpe, & \qquad & \myvec{J}=\vxm\kreu\vpm+\vxe\kreu\vpe, \end{eqnarray}
where $\vxm, \vpm$ and $\vxe, \vpe$ denote the canonical position
 and momentum variables of the particles with labels 1, 2.

In the following we will also need a slight generalization of an infinitesimal canonical
transformation. Let $g$ be the generator, that is a smooth function
on phase space. Define by \begin{subequations}\label{eq:howtoApplyGenerator}\begin{eqnarray}
T_{g}f & \assign & \exp\left(\PB{\cdot}{g}\right)f \\
 & = &
 f+\PB{f}{g}+\frac{1}{2}\PB{\PB{f}{g}}{g}+\frac{1}{6}\PB{\PB{\PB{f}{g}}{g}}{g}+
 \dots\end{eqnarray}\end{subequations}
an operator on phase-space functions $f$. It is easy to check that $T_{g}$
(i), preserves Poisson brackets and (ii), does not modify Kronecker deltas, and
therefore is a (finite) canonical transformation. For our purposes,
the generator will always be at least of order $\clight^{-2}$, and
we are only interested in a weakly relativistic portion. Cutting
off after the appropriate PN level, only finitely many
brackets remain in the above formula, and $T_{g}$ thereby turns into
a generalized infinitesimal canonical transformation.

\section{Center-of-mass and rest-frame canonical coordinates}

The well-known canonical coordinates adapted to the Newtonian problem obey the
usual canonical Poisson bracket relations exactly without reference
to any PN cutoff:\begin{subequations}\label{eq:NewtonJacobikoordinatenZwei}\begin{eqnarray}
\xz^{\op{N}} & = & \frac{\mm}{\mme}\vxm+\frac{\me}{\mme}\vxe,\\
\pz^{\op{N}} & = & \vpm+\vpe,\\
\xs^{\op{N}} & = & \vxm-\vxe,\\
\ps^{\op{N}} & = & \frac{\me}{\mme}\vpm-\frac{\mm}{\mme}\vpe, \end{eqnarray}\end{subequations}
with 
\begin{equation}
\PB{X^{\op{N} i}}{P^{\op{N}j}}=\delta^{ij},\quad \PB{x^{\op{N} i}}{p^{\op{N}j}}=\delta^{ij}, \quad 
\mbox{zero otherwise}.\end{equation}

A contact transformation will take ${\bf X}^{\rm N}$ to the correct Newton-Wigner center-of-mass coordinate $\xz$. In the spirit of Ref. \citep{BakTho53}, we impose\begin{subequations}\label{eq:BakTho53-Bedingungen}\begin{eqnarray}
T_{g}\xz^{\op{N}} & = & \xz, \\
\PB{\pz}{g} & = & 0,\\
\PB{\myvec{J}}{g} & = & 0, \\
g & = & O\left(\clight^{-2}\right) \end{eqnarray}\end{subequations}
on the generating function $g$ and find the new coordinates as \begin{subequations}\label{eq:explicitCoordinateTransform}\begin{eqnarray}
\xz & = & T_{g}\xz^{\op{N}}, \\
\pz & = & T_{g}\pz^{\op{N}},  \\
\xs & = & T_{g}\xs^{\op{N}}, \\
\ps & = & T_{g}\ps^{\op{N}}. \end{eqnarray}\end{subequations}
 These conditions have a number of desirable consequences. The new
set of fundamental canonical coordinates includes the center-of-mass
linear momentum and position vectors $\pz$ and $\xz$, respectively. Poisson bracket
relations similar to the particle variables are satisfied, no constraints
are necessary. The orbital angular momentum is realized simply as $\myvec{J}=\xz\kreu\pz+\xs\kreu\ps$.
They also entail that the center-of-energy vector $\myvec{G}=\myvec{G}\left(\xz,\pz,\myvec{J},M\right)$
depends only on the motion of the system as a whole, while the invariant
mass $M=M\left(\xs,\ps\right)$ is a function of the internal dynamics
only. The latter is important because $M$ generates the internal
part of the equations of motion: for any observable $f$, we have
$\PB{f}{H}=\clight^{2}MH^{-1}\PB{f}{\clight^{2}M}+\frac{1}{2}\clight^{2}H^{-1}\PB{f}{\pzpz}$.
Since $T_{g}$ is a canonical transformation, the Poincar\'e algebra is satisfied in the new variables.

The generator of the canonical transformation we calculated by the method
of undetermined coefficients \citep{DJS00}. All algebraic manipulations
were performed with the aid of \noun{Mathematica} and \noun{xTensor} \citep{MarGar07}. Groebner basis methods were used to account for the non-uniqueness resulting from the following vector identity: \begin{equation}
\parenmacro{\myvec{v}_{1}\cdot\myvec{v}_{2}\kreu\myvec{v}_{3}}\myvec{v}_{4}=\parenmacro{\myvec{v}_{4}\cdot\myvec{v}_{2}\kreu\myvec{v}_{3}}\myvec{v}_{1}+\parenmacro{\myvec{v}_{1}\cdot\myvec{v}_{4}\kreu\myvec{v}_{3}}\myvec{v}_{2}+\parenmacro{\myvec{v}_{1}\cdot\myvec{v}_{2}\kreu\myvec{v}_{4}}\myvec{v}_{3}.\end{equation}

\section{Results for two point masses without spin}

It is most convenient to introduce rescaled variables in the
form\begin{subequations}\label{eq:GeometricalUnits}\begin{equation}
\xsprime=\frac{\xs}{G\muz},\quad\psprime=\frac{\ps}{\mus},\quad\hat{t}=\frac{t}{G\muz},\quad\hat{M}=\frac{M-\muz}{\mus},\end{equation}
\begin{equation}
\rsprime=\left\vert \xsprime\right\vert ,\quad\nsprime=\xsprime/\rsprime
= \xs/\rs  = \ns,\end{equation}
\begin{equation}
\mus=\frac{\mm\me}{\mme},\quad\muz=\mme,\quad\nu=\frac{\mus}{\muz}. \end{equation}\end{subequations}
From the structure of our Hamiltonian in relation to the rest-mass energy it is clear that our 
reduced rest mass has to read \citep[Eqs. 3.2-3.6]{DJS00c}, \citep{DJS00b}, \citep{DJS01dimReg},

\begin{eqnarray}
\hat{M} & = & \frac{1}{2}\pspsprime-\frac{1}{\rsprime} + \clight^{-2}\lcur\frac{3\nu\
-1}{8}\left(\pspsprime\right)^{2}-\frac{\nu+3}{2}\frac{\pspsprime}{\rsprime}-\frac{\nu}{2}\frac{\nspsprime^{2}}{\rsprime}+\frac{1}{2}\frac{1}{\rsprime^{2}}\rcur\\
 & + & \clight^{-4}\lcur\frac{1}{16}\left(5\nu^{2}-5\nu\
+1\right)\left(\pspsprime\right)^{3}-\frac{3\
\nu^{2}+20\nu-5}{8}\frac{\left(\pspsprime\right)^{2}}{\rsprime}-\frac{\nu\
^{2}}{4}\frac{\nspsprime^{2}\pspsprime}{\rsprime} \nonumber\\
 &  & -\frac{3\nu^{2}}{8}\frac{\nspsprime^{4}}{\rsprime}+\left(4\nu\
+\frac{5}{2}\right)\frac{\pspsprime}{\rsprime^{2}}+\frac{3\nu}{2}\frac{\nspsprime^{2}}{\rsprime^{2}}-\frac{3\nu+1}{4}\frac{1}{\rsprime^{3}}\rcur \nonumber\\
 & + & \clight^{-6}\lcur\frac{5}{128}\
\left(7\nu^{3}-14\nu^{2}+7\nu-1\right)\left(\pspsprime\right)^{4}+\frac{1}{16}\left(-5\nu^{3}-53\nu^{2}+42\nu\
-7\right)\frac{\left(\pspsprime\right)^{3}}{\rsprime} \nonumber\\
 &  & +\frac{(2-3\nu)\nu^{2}}{16}\frac{\nspsprime^{2}\
\left(\pspsprime\right)^{2}}{\rsprime}-\frac{3(\nu-1)\nu\
^{2}}{16}\frac{\nspsprime^{4}\pspsprime}{\rsprime}-\frac{5\nu\
^{3}}{16}\frac{\nspsprime^{6}}{\rsprime} \nonumber\\
 &  & +\frac{109\nu^{2}+136\nu-27}{16}\
\frac{\left(\pspsprime\right)^{2}}{\rsprime^{2}}+\frac{\nu\left(30\nu+17\right)}{16}\frac{\nspsprime^{2}\
\pspsprime}{\rsprime^{2}} \nonumber\\
 &  & +\frac{\nu(43\nu+5)}{12}\
\frac{\nspsprime^{4}}{\rsprime^{2}}-\frac{552\nu^{2}+\left(1340-3\pi^{2}\right)\nu\
+600}{192}\
\frac{\pspsprime}{\rsprime^{3}} \nonumber\\
 &  & -\frac{\nu\left(112\nu+3\pi^{2}+340\right)}{64}\frac{\nspsprime^{2}}{\rsprime^{3}}+\frac{\left(872-63\pi^{2}\right)\nu+12}{96}\frac{1}{\rsprime^{4}}\rcur+O\left(\clight^{-8}\right)\,, \nonumber\end{eqnarray}
and the non-center-of-mass Hamiltonian has to be given by 

\begin{equation}
H=\sqrt{\clight^{4}\left(\mus\hat{M}+\muz\right)^{2}+\clight^{2}\pzpz}.\end{equation}

The generating function $g_{\op{point}}$ 
as detailed by  Eqs. (\ref{eq:howtoApplyGenerator}) then turns out to read 
(note we are working with four different mass expressions $\mm,\me,\muz,\mus$
for two different point masses to cut down on formula length), where $\rme=\left\vert \vxm-\vxe\right\vert$, $\nme=\left( \vxm-\vxe\right)/\rme$,\begin{equation}
g_{\op{point}}=\frac{1}{2}\left[g_{\op{1pN}}+g_{\op{2pN}}+g_{\op{3pN}}^{\op{0G}}+g_{\op{3pN}}^{\op{1G}}+g_{\op{3pN}}^{\op{2G}}+g_{\op{3pN}}^{\op{3G}}+\left(\pqm\leftrightarrow\pqe\right)\right],\end{equation}
\begin{subequations}\begin{eqnarray}
g_{\op{1pN}} & = & \clight^{-2}\lcur\lbra\frac{\me^{2}}{\mm\muz^{3}}\nmepm\pmpm\rme+\frac{\me^{2}}{\mm\muz^{3}}\nmepe\pmpm\rme \nonumber\\
 &  & -\frac{2\me-\muz}{\muz^{3}}\nmepm\pmpe\rme\rbra-G\frac{\mus(2\me-\muz)}{\muz}\nmepm\rcur,\end{eqnarray}
\begin{eqnarray}
g_{\op{2pN}} & = & \clight^{-4}\lcur\lbra-\frac{\me^{2}(\me+2\mm)}{4\mm^{3}\muz^{4}}\nmepm\left(\pmpm\right)^{2}\
\rme-\frac{\me^{2}(\me+2\mm)}{4\mm^{3}\muz^{4}}\nmepe\
\left(\pmpm\right)^{2}\rme \nonumber\\
 &  & +\frac{2\me+\mus-\muz}{2\mm^{2}\
\muz^{3}}\nmepm\pmpe\pmpm\rme+\frac{2\me+\mus-\muz}{2\mm^{2}\
\muz^{3}}\nmepe\pmpe\pmpm\rme \nonumber\\
 &  & -\frac{2\me-\muz}{2\mus\
\muz^{4}}\nmepm\pmpe^{2}\rme-\frac{2\me-\muz}{4\mus\
\muz^{4}}\nmepm\pepe\pmpm\rme\rbra \nonumber\\
 &  & -G\lbra\frac{\me^{2}(\me-\mm)}{4\muz^{4}}\nmepm^{3}+\frac{\mm(-5\me+6\
\mus+\muz)}{4\muz^{3}}\nmepe\nmepm^{2} \nonumber\\
 &  & +\frac{2\
\mus(3\mus-8\muz)-5\me(\mus-2\muz)}{4\mm\muz^{2}}\nmepm\
\pmpm \nonumber\\
 &  & +\frac{6\mus^{2}-5\me\mus-16\muz\mus+5\muz^{2}+5\
\me\muz}{4\mm\muz^{2}}\nmepe\pmpm \nonumber\\
 &  & +\frac{6\muz^{2}-17\
\me\muz-3\mus\muz+6\me\mus}{2\muz^{3}}\nmepm\pmpe\rbra \nonumber\\
 &  & +G^{2}\frac{\mus(2\mus-\muz)(\muz-2\me)}{2\
\muz}\frac{\nmepm}{\rme}\rcur,\end{eqnarray}
\begin{eqnarray}
g_{\op{3pN}}^{\op{0G}} & = & \clight^{-6}\lcur\frac{\me\left(-\mus^{2}+6\muz^{2}\right)+2\mus\left(\mus^{2}+5\
\muz\mus-3\muz^{2}\right)}{48\mm^{5}\muz^{4}}\nmepm\left(\pmpm\right)^{3}\
\rme \nonumber\\
 &  & +\frac{\me\left(-\mus^{2}+6\muz^{2}\right)+2\mus\
\left(\mus^{2}+5\muz\mus-3\muz^{2}\right)}{48\mm^{5}\
\muz^{4}}\nmepe\left(\pmpm\right)^{3}\rme \nonumber\\
 &  & -\frac{\muz\left(3\
\mus^{2}+16\muz\mus-8\muz^{2}\right)+\me\left(-6\mus^{2}-20\muz\mus+17\muz^{2}\right)}{24\mm^{4}\muz^{5}}\nmepm\pmpe\left(\pmpm\right)^{2}\
\rme \nonumber\\
 &  & -\frac{\muz\left(3\mus^{2}+16\muz\mus-8\
\muz^{2}\right)+\me\left(-6\mus^{2}-20\muz\mus+17\muz^{2}\right)}{24\
\mm^{4}\muz^{5}}\nmepe\pmpe\left(\pmpm\right)^{2}\rme \nonumber\\
 &  & \
+\frac{\me\left(-6\mus^{2}-14\muz\mus+13\muz^{2}\right)+\muz\
\left(3\mus^{2}+7\muz\mus-6\muz^{2}\right)}{12\me\mm^{3}\
\muz^{5}}\nmepm\pmpe^{2}\pmpm\rme \nonumber\\
 &  & -\frac{\me\
\left(6\mus^{2}+14\muz\mus-13\muz^{2}\right)+\muz\left(-3\mus^{2}-7\muz\mus+6\muz^{2}\right)}{12\me\mm^{3}\muz^{5}}\nmepe\pmpe^{2}\
\pmpm\rme \nonumber\\
 &  & +\frac{(2\me-\muz)\left(\mus^{2}+2\
\muz\mus-2\muz^{2}\right)}{6\mus^{2}\muz^{7}}\nmepm\
\pmpe^{3}\rme \nonumber\\
 &  & -\frac{2\me\left(3\mus^{2}+7\muz\
\mus-6\muz^{2}\right)+\muz\left(-3\mus^{2}-16\muz\mus+7\
\muz^{2}\right)}{48\me\mm^{3}\muz^{5}}\nmepm\pepe\left(\pmpm\right)^{2}\
\rme \nonumber\\
 &  & -\frac{2\me\left(3\mus^{2}+7\muz\mus-6\
\muz^{2}\right)+\muz\left(-3\mus^{2}-16\muz\mus+7\muz^{2}\right)}{48\
\me\mm^{3}\muz^{5}}\nmepe\pepe\left(\pmpm\right)^{2}\rme \nonumber\\
 &  & +\frac{(2\me-\muz)\left(3\mus^{2}+6\muz\mus-5\
\muz^{2}\right)}{12\mus^{2}\muz^{7}}\nmepm\pepe\pmpe\pmpm\
\rme\rcur,\end{eqnarray}
\begin{eqnarray}
g_{\op{3pN}}^{\op{1G}} & = & \clight^{-6}G\lcur\frac{\me^{2}(\mm-\me)}{16\muz^{6}}\nmepm^{5}  +\frac{10\
\mus^{2}-\me\mus-16\muz\mus+\muz^{2}+5\me\muz}{16\mm\
\muz^{4}}\nmepe\nmepm^{4} \nonumber\\
 &  & +\frac{10\
\mus^{2}+3\me\mus-26\muz\mus+8\muz^{2}-5\me\muz}{8\mm\
\muz^{4}}\nmepe^{2}\nmepm^{3} \nonumber\\
 &  & \
+\frac{\me^{2}(10\me+4\mus-3\muz)}{48\mm\
\muz^{5}}\nmepm^{3}\pmpm \nonumber\\
 &  & -\frac{4\mus^{2}-13\
\muz\mus+\muz^{2}+6\me(\mus+\muz)}{16\mm\muz^{4}}\nmepe\
\nmepm^{2}\pmpm \nonumber\\
 &  & +\frac{\muz\left(6\mus^{2}+9\
\muz\mus-4\muz^{2}\right)+\me\left(-4\mus^{2}+15\muz\mus+5\
\muz^{2}\right)}{16\mus\muz^{5}}\nmepe^{2}\nmepm\
\pmpm \nonumber\\
 &  & -\frac{\me\left(27\mus^{2}-32\muz\mus-54\
\muz^{2}\right)+2\mus\left(\mus^{2}-9\muz\mus+43\muz^{2}\right)}{48\
\mm^{3}\muz^{3}}\nmepm\left(\pmpm\right)^{2} \nonumber\\
 &  & -\frac{\me\
\left(27\mus^{2}-20\muz\mus-42\muz^{2}\right)+2\left(\mus^{3}-9\muz\
\mus^{2}+43\muz^{2}\mus-6\muz^{3}\right)}{48\mm^{3}\muz^{3}}\nmepe\
\left(\pmpm\right)^{2} \nonumber\\
 &  & +\frac{-4\me\mus+2\muz\mus+3\me\
\muz}{24\muz^{5}}\nmepm^{3}\pmpe \nonumber\\
 &  & \
-\frac{\muz\left(-2\mus^{2}+7\muz\mus-6\muz^{2}\right)+\me\left(4\
\mus^{2}-15\muz\mus+18\muz^{2}\right)}{8\mus\muz^{5}}\nmepe\
\nmepm^{2}\pmpe \nonumber\\
 &  & -\frac{\muz\left(-30\mus^{2}+84\
\muz\mus-47\muz^{2}\right)+\me\left(4\mus^{2}-38\muz\mus+101\
\muz^{2}\right)}{24\mm^{2}\muz^{4}}\nmepm\pmpe\pmpm \nonumber\\
 &  & -\frac{-4\mus^{3}+38\muz\mus^{2}-101\muz^{2}\mus-6\muz^{3}+\me\
\left(-26\mus^{2}+22\muz\mus+60\muz^{2}\right)}{24\me\mm^{2}\
\muz^{3}}\nmepe\pmpe\pmpm \nonumber\\
 &  & -\frac{\me\left(-2\mus^{2}+14\
\muz\mus-55\muz^{2}\right)+\muz\left(\mus^{2}-13\muz\mus+29\
\muz^{2}\right)}{12\mus\muz^{5}}\nmepm\pmpe^{2} \nonumber\\
 &  & \
-\frac{\me\left(4\mus^{2}-11\muz\mus-43\muz^{2}\right)+\muz\
\left(2\mus^{2}+13\muz\mus+13\muz^{2}\right)}{48\mus\
\muz^{5}}\nmepm^{3}\pepe \nonumber\\
 &  & -\frac{\me\left(-4\
\mus^{2}+60\muz\mus+2\muz^{2}\right)+\muz\left(2\mus^{2}-42\muz\
\mus+17\muz^{2}\right)}{48\mus\muz^{5}}\nmepm\pepe\pmpm\rcur,\end{eqnarray}
\begin{eqnarray}
g_{\op{3pN}}^{\op{2G}} & = & \clight^{-6}G^{2}\lcur\frac{\me\left(24\mus^{2}-20\muz\mus+5\muz^{2}+2\me\
(\muz-6\mus)\right)}{24\muz^{3}}\frac{\nmepm^{3}}{\rme}\\
 &  & \
-\frac{\muz\left(12\mus^{2}-6\muz\mus-5\muz^{2}\right)+\me\
\left(-72\mus^{2}+48\muz\mus+3\muz^{2}\right)}{24\
\muz^{3}}\frac{\nmepe\nmepm^{2}}{\rme} \nonumber\\
 &  & +\frac{\me\
\left(38\mus^{2}+19\muz\mus+90\muz^{2}\right)-2\mus\left(22\
\mus^{2}-33\muz\mus+68\muz^{2}\right)}{24\mm\
\muz^{2}}\frac{\nmepm\pmpm}{\rme} \nonumber\\
 &  & +\frac{-44\
\mus^{3}+66\muz\mus^{2}-166\muz^{2}\mus+65\muz^{3}+\me\left(38\
\mus^{2}+37\muz\mus+25\muz^{2}\right)}{24\mm\
\muz^{2}}\frac{\nmepe\pmpm}{\rme} \nonumber\\
 &  & -\frac{\muz\
\left(-44\mus^{2}+62\muz\mus-79\muz^{2}\right)+4\me\left(22\
\mus^{2}-40\muz\mus+69\muz^{2}\right)}{24\muz^{3}}\frac{\nmepm\
\pmpe}{\rme}\rcur, \nonumber\end{eqnarray}
\begin{eqnarray}
g_{\op{3pN}}^{\op{3G}} & = & \clight^{-6}G^{3}\frac{\mus(\muz-2\me)\left(4\mus^{2}-\muz\mus+\muz^{2}\right)}{4\muz}\frac{\nmepm}{\rme^{2}}.\end{eqnarray}\end{subequations}

\section{Results for two particles with spin}

In the case of two spinning particles, the Hamiltonian and the center-of-energy vector 
are only known up to the formal 2PN order. Here, we are
counting formally, i.e. $\sm$ has the same units as an orbital angular momentum,
without any reference to its magnitude for maximally rotating black holes. 
Since the expression for $H$ is quite lenghty, we will not repeat
here the formulas for the Hamiltonians given in
\citep{SteiSchaHerg08,HergScha08a,HergScha08b,SteiHergScha08}. However, the total Hamiltonian
may be abbreviated as follows, \citep{Schae09Orleans},
\begin{eqnarray}
H &=& H_{N}+H_{1PN}+H_{2PN}+H_{3PN}  \nonumber \\[1ex]
&+& H_{SO}^{\rm LO}+H^{\rm LO}_{S_{1}S_{2}} + H^{\rm LO}_{S_{1}^2}+
H^{\rm LO}_{S_{2}^2} \nonumber \\[1ex]
&+& H_{SO}^{\rm NLO}+H^{\rm NLO}_{S_{1}S_{2}} + H^{\rm NLO}_{S_{1}^2}+ H^{\rm NLO}_{S_{2}^2} \nonumber \\ [1ex]
 &+& H_{p_{1}S_{2}^3}+H_{p_{2}S_{1}^3} + H_{p_{1}S_{1}^3}+H_{p_{2}S_{2}^3} \nonumber \\[1ex]
&+& H_{p_{1}S_{1}S_{2}^2} + H_{p_{2}S_{2}S_{1}^2} + H_{p_{1}S_{2}S_{1}^2} + H_{p_{2}S_{1}S_{2}^2} \nonumber \\[1ex]
&+&  H_{S_{1}^2S_{2}^2} + H_{S_{1} S_{2}^3} + H_{S_{2}S_{1}^3}\,,
\end{eqnarray}
where LO and NLO respectively denote leading and next-to-leading
order coupling and $SO$ spin-orbit coupling. The other Hamiltonians with
spin are leading order ones. The related expression for $\hat{M}$ is given by

\begin{equation}
\hat{M}_{\op{2pN\, spin}}^{\op{total}}=\hat{M}_{\op{point}}+\frac{1}{2}\left[\hat{M}_{\op{1pN}}^{\op{s}}+\hat{M}_{\op{2pN}}^{\op{s}}+\left(1\leftrightarrow2\right)\right]+O\left(\clight^{-6}\right)\,.\end{equation}
Here,  {}``$1\leftrightarrow2$'' maps $\ps\leftrightarrow-\ps$,
$\xs\leftrightarrow-\xs$, $\scm\leftrightarrow\sce$, and $m_1 \leftrightarrow m_2$. In an abuse of notation, we will
take $\hat{M}_{\op{point}}$ to signal the same dependency on the
variables $\xs$, $\ps$ as in the spinless case. Since these have
a different meaning here, $\hat{M}_{\op{point}}$ is not the same
phase-space function as $\hat{M}$  in the previous section. Introducing \begin{equation}
\scmprime=\frac{\scm}{G \mm},\quad\sceprime=\frac{\sce}{G \me},\end{equation}
where 
\begin{subequations}\begin{eqnarray}
\scm & = & T_{g_{\op{spin}}^{\op{total}}}\sm,\\
\sce & = & T_{g_{\op{spin}}^{\op{total}}}\se,\end{eqnarray}\end{subequations}
we get
\begin{subequations}\begin{eqnarray}
{\hat M}_{\op{1pN}}^{\op{s}} & = & \clight^{-2}
 \lcur \frac{4 \muz-\me}{\muz^2}\frac{\epsnspsscmprime}{\rs^2} +\frac{3}{\muz^2}\frac{\nsscmprime^2}{\rsprime^3}\nonumber\\
 & & +\frac{3}{\muz^2}\frac{\nssceprime \nsscmprime}{\rs^3} -\frac{1}{\muz^2}\frac{\scmscmprime}{\rs^3}-\frac{1}{\muz^2}\frac{\scmsceprime}{\rs^3}
 \rcur,\end{eqnarray}

\begin{eqnarray}
{\hat M}_{\op{2pN}}^{\op{s}} & = & \clight^{-4}\lcur \frac{2 (\me-2 (\mus+3 \muz))}{\muz^2}\frac{\epsnspsscmprime}{\rsprime^3}+\frac{3 \me \mm (\me+2 \mm)}{2 \muz^4}\frac{\epsnspsscmprime \nspsprime^2}{\rsprime^2}\nonumber\\
 &  &  -\frac{3 \me \mus-19 \muz \mus+5 \me \muz}{4 \muz^3}\frac{\epsnspsscmprime \pspsprime}{\rsprime^2} -\frac{5 \me+9 \muz}{\muz^3}\frac{\nsscmprime^2}{\rsprime^4}\nonumber\\
 & & +\frac{15 \mus}{2 \muz^3}\frac{\nspsprime^2 \nsscmprime^2}{\rsprime^3} -\frac{12}{\muz^2}\frac{\nssceprime \nsscmprime}{\rsprime^4}\nonumber\\
 & & +\frac{15 \mus}{2 \muz^3}\frac{\nspsprime^2 \nssceprime \nsscmprime}{\rsprime^3} +\frac{3 (-7 \me+\mus+6 \muz)}{4 \muz^3}\frac{\nsscmprime^2 \pspsprime}{\rsprime^3}\nonumber\\
  &  & +\frac{9 \mus+6 \muz}{4 \muz^3}\frac{\nssceprime \nsscmprime \pspsprime}{\rsprime^3} -\frac{3 (\me+\mus)}{2 \muz^3}\frac{\nspsprime \nsscmprime \psscmprime}{\rsprime^3} \nonumber\\
 & & +\frac{3 (2 \me-3 \mus-4 \muz)}{2 \muz^3}\frac{\nspsprime \nssceprime \psscmprime}{\rsprime^3} +\frac{\me^2}{2 \muz^4}\frac{\psscmprime^2}{\rsprime^3}\nonumber\\
  &  & +\frac{\mus+3 \muz}{2 \muz^3}\frac{\pssceprime \psscmprime}{\rsprime^3} +\frac{\me+5 \muz}{\muz^3}\frac{\scmscmprime}{\rsprime^4}+\frac{3 (\me-3 \mus)}{4 \muz^3}\frac{\nspsprime^2 \scmscmprime}{\rsprime^3}\nonumber\\
  &  & -\frac{3 \mm}{2 \muz^3}\frac{\pspsprime \scmscmprime}{\rsprime^3} +\frac{6}{\muz^2}\frac{\scmsceprime}{\rsprime^4}-\frac{3 (\mus-4 \muz)}{4 \muz^3}\frac{\nspsprime^2 \scmsceprime}{\rsprime^3}\nonumber\\
  &  & -\frac{2 \mus+3 \muz}{2 \muz^3}\frac{\pspsprime \scmsceprime}{\rsprime^3} -\frac{5 (\me+4 \mm)}{2 \muz^4}\frac{\epsnspsscmprime \nsscmprime^2}{\rsprime^4}\nonumber\\
  &  &  +\frac{\me+4 \mm}{2 \muz^4}\frac{\epsnspsscmprime \scmscmprime}{\rsprime^4} +\frac{15 (\me-3 \muz)}{2 \muz^4}\frac{\epsnspssceprime \nsscmprime^2}{\rsprime^4}\nonumber\\
  &  & -\frac{3 (\me-3 \muz)}{2 \muz^4}\frac{\epsnspssceprime \scmscmprime}{\rsprime^4} -\frac{15 \me}{\muz^4}\frac{\epsnsscmsceprime \nspsprime \nsscmprime}{\rsprime^4}\nonumber\\
  &  & +\frac{3 \me}{\muz^4}\frac{\epsnsscmsceprime \psscmprime}{\rsprime^4}+\frac{6 \me-9 \muz}{\muz^4}\frac{\epspsscmsceprime \nsscmprime}{\rsprime^4}\nonumber\\
&  & -\frac{35}{\muz^4}\frac{\nssceprime \nsscmprime^3}{\rsprime^5} -\frac{105}{4 \muz^4}\frac{\nssceprime^2 \nsscmprime^2}{\rsprime^5}\nonumber\\
&  & +\frac{15}{\muz^4}\frac{\nssceprime \nsscmprime \scmscmprime}{\rsprime^5}+\frac{15}{2 \muz^4}\frac{\nssceprime^2 \scmscmprime}{\rsprime^5}\nonumber\\
&  & +\frac{15}{\muz^4}\frac{\nsscmprime^2 \scmsceprime}{\rsprime^5}+\frac{15}{\muz^4}\frac{\nssceprime \nsscmprime \scmsceprime}{\rsprime^5}\nonumber\\
&  & -\frac{3}{\muz^4}\frac{\scmsceprime \scmscmprime}{\rsprime^5}+\frac{3}{4 \muz^4}\frac{\scmsceprime^2}{\rsprime^5} -\frac{3}{4 \muz^4}\frac{\scesceprime \scmscmprime}{\rsprime^5}
 \rcur.  \end{eqnarray}\end{subequations}

Again, we have determined
a generating function such that the binary's invariant mass takes
the same form in the transformed variables as in the center-of-mass frame. 
However, the Poincar\'e group generators $H$, $\myvec{G}$,
$\myvec{J}$ and therefore also $\xz$ differ from the point-particle
case, particularly $\myvec{J}=\vxm\kreu\vpm+\vxe\kreu\vpe + \sm + \se$,
where additionally
$\PB{S^{i}_a}{S^{j}_b}=\epsilon^{ijk}S^{k}\delta_{ab}$  holds.

The generating function $g_{\op{spin}}^{\op{total}}$ reads, 
\begin{equation}
g_{\op{spin}}^{\op{total}}=g_{\op{point}}+\frac{1}{2}\left[g_{\op{spin}}+\left(\pqm\leftrightarrow\pqe\right)\right],\end{equation}
\begin{eqnarray}
g_{\op{spin}} & = & -\clight^{-2}\frac{1}{\mm\muz}\epspmpeSm+\clight^{-4}\lcur\lbra\frac{\mus+\muz}{4\mm^{3}\muz^{2}}\epspmpeSm\pmpm \nonumber\\
 &  & +\frac{2\me-3\muz}{4\mm^{2}\muz^{3}}\epspmpeSm\pmpe-\frac{\me+\muz}{4\mus\muz^{4}}\epspmpeSe\pmpm\rbra \nonumber\\
 &  & +G\lbra\frac{3(\me-2\mus)}{2\mm\muz}\frac{\nmepm\nmeSm^{2}}{\rme^{2}}+\frac{6\me-3\muz}{\muz^{2}}\frac{\nmepm\nmeSe\nmeSm}{\rme^{2}} \nonumber\\
 &  & +\frac{3(\me+2\mus-\muz)}{2\me\muz}\frac{\nmepm\nmeSe^{2}}{\rme^{2}}-\frac{4\me}{\mm\muz}\frac{\nmeSm\pmSm}{\rme^{2}} \nonumber\\
 &  & +\frac{1}{\muz}\frac{\nmeSe\pmSm}{\rme^{2}}-\frac{1}{\muz}\frac{\nmeSm\pmSe}{\rme^{2}} \nonumber\\
 &  & +\frac{4\mm}{\me\muz}\frac{\nmeSe\pmSe}{\rme^{2}}+\frac{3\me+2\mus}{2\mm\muz}\frac{\nmepm\SmSm}{\rme^{2}} \nonumber\\
 &  & +\frac{\muz-2\me}{\muz^{2}}\frac{\nmepm\SmSe}{\rme^{2}}-\frac{\mm(2\me+3\muz)}{2\me\muz^{2}}\frac{\nmepm\SeSe}{\rme^{2}} \nonumber\\
 &  & +\frac{2\me(\mus+5\muz)-17\mus\muz}{4\mm\muz^{2}}\frac{\epsnmepmSm\nmepm}{\rme} \nonumber\\
 &  & +\frac{4\muz^{2}+6\me\muz-17\mus\muz+2\me\mus}{4\mm\muz^{2}}\frac{\epsnmepmSm\nmepe}{\rme} \nonumber\\
 &  & +\frac{19\me-2\mus-7\muz}{4\muz^{2}}\frac{\epsnmepmSe\nmepm}{\rme} \nonumber\\
 &  & -\frac{19\me+2\mus-8\muz}{4\muz^{2}}\frac{\epsnmepeSm\nmepm}{\rme} \nonumber\\
 &  & -\frac{\me+2\mus-6\muz}{2\mm\muz}\frac{\epspmpeSm}{\rme}\rbra\rcur.\end{eqnarray}

\section{Conclusions}

We have derived the center-of-mass and rest-frame coordinate representation
of the presently known binary Hamiltonians with spinning components. The
result is given by Eq. (17) and Eqs. (25). 
Additionally, and even more important, all the transformations from the original canonical
coordinates to the new ones are presented in explicit form in the Eqs. (19),
(20), (26), and (27). This shows full consistency of our general relativistic Hamiltonians
having to be of the form (18) with reduced rest-mass only depending on rest-frame coordinates.  
Thus, our curved spacetimes generated through interacting spinning
bodies perfectly respect the asymptotic Lorentz invariance as they should.

\acknowledgments The authors thank Jan Steinhoff for useful discussions.
This work is supported by the Deutsche Forschungsgemeinschaft
(DFG) through SFB/TR7 {}``Gravitational Wave Astronomy'' and the DFG 
project ``Kr\"ummungsma{\ss}e f\"ur Fraktale und Anwendung in der Mustererkennung''.

\appendix

\section{Generators given in center-of-mass coordinates}

For the convenience of the reader, we also give the resulting generating functions
in center-of-mass coordinates. The canonical transformation generated
by $-g_{\op{point}}$ or $-g_{\op{spin}}^{\op{total}}$ respectively
maps the center-of-mass coordinates onto their Newtonian counterparts:

\begin{equation}
-g_{\op{point}}=h_{\op{1pN}}+h_{\op{2pN}}+h_{\op{3pN}}^{\op{0G}}+h_{\op{3pN}}^{\op{1G}}+h_{\op{3pN}}^{\op{2G}}+h_{\op{3pN}}^{\op{3G}}+O\left(\clight^{-8}\right),\end{equation}

\begin{equation}
-g_{\op{spin}}^{\op{total}}=h_{\op{1pN}}+h_{\op{2pN}}+\frac{1}{2}\left[h_{\op{spin}}+\left(1\leftrightarrow2\right)\right]+O\left(\clight^{-6}\right),\end{equation}

\begin{subequations}\begin{eqnarray}
h_{\op{1pN}} & = & \clight^{-2}\lcur-\frac{1}{2\muz^{2}}\nspz\pzps\rs+\frac{\muz-2\me}{2\mus\muz^{2}}\nspz\psps\rs\rcur \nonumber\\
& & +\clight^{-2}G\mus\left(\frac{\me}{\muz}-\frac{1}{2}\right)\nspz,\end{eqnarray}
\begin{eqnarray}
h_{\op{2pN}} & = & \clight^{-4}\lcur\frac{1}{4\muz^{4}}\nspz\pzps\pzpz\rs+\frac{\me-\mm}{4\mus\muz^{4}}\nspz\pzps^{2}\rs+\frac{\me-\mm}{8\mus\muz^{4}}\nspz\psps\pzpz\rs \nonumber\\
 &  & +\frac{\muz-2\mus}{4\mus^{2}\muz^{3}}\nspz\psps\pzps\rs-\frac{\muz-2\me}{8\mus^{3}\muz^{2}}\nspz\left(\psps\right)^{2}\rs\rcur \nonumber\\
 &  & +\clight^{-4}G\lcur\frac{\mus(\muz-2\me)}{8\muz^{3}}\nspz^{3}-\frac{\muz-4\mus}{8\muz^{2}}\nsps\nspz^{2}+\frac{\me-\mm}{4\muz^{2}}\nsps^{2}\nspz \nonumber\\
 &  & +\frac{\mus(\muz-2\me)}{8\muz^{3}}\nspz\pzpz-\frac{5}{8\muz}\nsps\pzpz+\frac{1}{\muz}\nspz\pzps \nonumber\\
 &  & +\frac{(2\mus-5\muz)(\muz-2\me)}{4\mus\muz^{2}}\nspz\psps\rcur-\clight^{-4}G^{2}\frac{\mus(2\mus-\muz)(\muz-2\me)}{4\muz}\frac{\nspz}{\rs},\end{eqnarray}
\begin{eqnarray}
h_{\op{3pN}}^{\op{0G}} & = & \clight^{-6}\lcur-\frac{1}{6\muz^{6}}\nspz\pzps\left(\pzpz\right)^{2}\rs+\frac{\muz-2\me}{4\mus\muz^{6}}\nspz\pzps^{2}\pzpz\rs \nonumber\\
 &  & +\frac{5\mus-2\muz}{12\mus^{2}\muz^{5}}\nspz\pzps^{3}\rs+\frac{7(\muz-2\me)}{96\mus\muz^{6}}\nspz\psps\left(\pzpz\right)^{2}\rs \nonumber\\
 &  & +\frac{8\mus-5\muz}{24\mus^{2}\muz^{5}}\nspz\psps\pzps\pzpz\rs-\frac{(4\mus-7\muz)(\muz-2\me)}{24\mus^{3}\muz^{5}}\nspz\psps\pzps^{2}\rs \nonumber\\
 &  & -\frac{(2\me-\muz)(4\mus+5\muz)}{96\mus^{3}\muz^{5}}\nspz\left(\psps\right)^{2}\pzpz\rs+\frac{10\mus-3\muz}{16\mus^{4}\muz^{3}}\nspz\left(\psps\right)^{2}\pzps\rs \nonumber\\
 &  & +\frac{\mm^{4}-\me^{4}}{16\mus^{5}\muz^{5}}\nspz\left(\psps\right)^{3}\rs\rcur,\end{eqnarray}
\begin{eqnarray}
h_{\op{3pN}}^{\op{1G}} & = & \clight^{-6}G\lcur\frac{\me(\me-\mm)\mm}{32\muz^{6}}\nspz^{5}+\frac{\muz-4\mus}{32\muz^{4}}\nsps\nspz^{4} \nonumber\\
 &  & +\frac{3(\muz-2\me)}{16\muz^{4}}\nsps^{2}\nspz^{3}-\frac{3(\muz-4\mus)}{16\mus\muz^{3}}\nsps^{3}\nspz^{2} \nonumber\\
 &  & -\frac{3(\mm-\me)}{16\mus\muz^{3}}\nsps^{4}\nspz+\frac{\me(\me-\mm)\mm}{12\muz^{6}}\nspz^{3}\pzpz \nonumber\\
 &  & +\frac{3\muz-8\mus}{32\muz^{4}}\nsps\nspz^{2}\pzpz-\frac{\left(2\me-\muz\right)\left(\mus+2\muz\right)}{16\mus\muz^{4}}\nsps^{2}\nspz\pzpz \nonumber\\
 &  & -\frac{3}{16\mus\muz^{2}}\nsps^{3}\pzpz+\frac{7\me\mm\left(\me-\mm\right)}{96\muz^{6}}\nspz\left(\pzpz\right)^{2} \nonumber\\
 &  & +\frac{1}{4\muz^{3}}\nsps\left(\pzpz\right)^{2}+\frac{5\left(\muz-4\mus\right)}{48\muz^{4}}\nspz^{3}\pzps \nonumber\\
 &  & +\frac{(2\mus-3\muz)(\muz-2\me)}{8\mus\muz^{4}}\nsps\nspz^{2}\pzps \nonumber\\
 &  & +\frac{1}{4\mus\muz^{2}}\nsps^{2}\nspz\pzps+\frac{4\mus-13\muz}{24\muz^{4}}\nspz\pzps\pzpz  \nonumber\\
 &  & +\frac{\me-\mm}{8\mus\muz^{3}}\nsps\pzps\pzpz-\frac{(8\mus-29\muz)(\muz-2\me)}{24\mus\muz^{4}}\nspz\pzps^{2}  \nonumber \\
 &  & -\frac{1}{8\mus\muz^{2}}\nsps\pzps^{2}-\frac{(4\mus-13\muz)(\muz-2\me)}{96\mus\muz^{4}}\nspz^{3}\psps  \nonumber\\
 &  & -\frac{1}{2\muz^{3}}\nsps\nspz^{2}\psps+\frac{\muz-2\me}{4\mus\muz^{3}}\nsps^{2}\nspz\psps  \nonumber\\
 &  & -\frac{(32\mus-17\muz)(\muz-2\me)}{96\mus\muz^{4}}\nspz\psps\pzpz  \nonumber\\
 &  & +\frac{1}{4\mus\muz^{2}}\nsps\psps\pzpz-\frac{8\mus^{2}-26\muz\mus+9\muz^{2}}{8\mus^{2}\muz^{3}}\nspz\psps\pzps  \nonumber\\
 &  & +\frac{9}{16}\left(\frac{1}{\me^{3}}-\frac{1}{\mm^{3}}\right)\nspz\left(\psps\right)^{2}\rcur,\end{eqnarray}
\begin{eqnarray}
h_{\op{3pN}}^{\op{2G}} & = & \clight^{-6}G^{2}\lcur\frac{\mus(6\mus-5\muz)(\muz-2\me)}{24\muz^{3}}\frac{\nspz^{3}}{\rs}+\left(\frac{\mus^{2}}{\muz^{2}}-\frac{3\mus}{4\muz}+\frac{5}{48}\right)\frac{\nsps\nspz^{2}}{\rs} \nonumber\\
 &  & +\frac{(\muz-2\me)(\muz-2\mus)}{4\muz^{2}}\frac{\nsps^{2}\nspz}{\rs}+\frac{\mus(7\mus-4\muz)(\muz-2\me)}{24\muz^{3}}\frac{\nspz\pzpz}{\rs} \nonumber\\
 &  & +\left(\frac{65}{48}-\frac{7\mus}{8\muz}\right)\frac{\nsps\pzpz}{\rs}+\left(\frac{\mus^{2}}{2\muz^{2}}-\frac{9\mus}{4\muz}-\frac{101}{48}\right)\frac{\nspz\pzps}{\rs} \nonumber\\
 &  & +\frac{\muz-2\me}{4\muz}\frac{\nsps\pzps}{\rs}-\frac{(2\me-\muz)\left(8\mus^{2}+5\muz\mus+30\muz^{2}\right)}{16\mus\muz^{2}}\frac{\nspz\psps}{\rs}\rcur,\end{eqnarray}
\begin{eqnarray}
h_{\op{3pN}}^{\op{3G}} & = & \clight^{-6}G^{3}\frac{\mus(2\me-\muz)\left(4\mus^{2}-\muz\mus+\muz^{2}\right)}{8\muz}\frac{\nspz}{\rs^{2}},\end{eqnarray}\end{subequations}

\begin{eqnarray}
h_{\op{spin}} & = & \clight^{-2}\frac{-1}{\mm\muz}\epspzpsscm+\clight^{-4}\lcur\frac{1}{4\mm\muz^{3}}\epspzpsscm\pzpz+\frac{1}{4\mm^{2}\muz^{2}}\epspzpsscm\pzps \nonumber \\ 
 &  & +\frac{\me+2\mm}{4\me\mm^{3}\muz}\epspzpsscm\psps\rcur+\clight^{-4}G\lcur-\frac{3(\me-2\mus)}{2\mm\muz}\frac{\nspz\nsscm^{2}}{\rs^{2}} \nonumber \\
 &  & +\frac{3(\muz-2\me)}{2\muz^{2}}\frac{\nspz\nssce\nsscm}{\rs^{2}}+\frac{4\me}{\mm\muz}\frac{\nsscm\pzscm}{\rs^{2}}-\frac{1}{\muz}\frac{\nssce\pzscm}{\rs^{2}} \nonumber \\
 &  & -\frac{3\me+2\mus}{2\mm\muz}\frac{\nspz\scmscm}{\rs^{2}}+\frac{\me-\mm}{2\muz^{2}}\frac{\nspz\scmsce}{\rs^{2}}-\frac{3\me+2\mus}{4\muz^{2}}\frac{\epsnspzscm\nspz}{\rs} \nonumber \\
 &  & +\frac{1}{\muz}\frac{\epsnspzscm\nsps}{\rs}-\frac{9\me+2\mus-4\muz}{2\mm\muz}\frac{\epsnspsscm\nspz}{\rs} \nonumber \\
 &  & -\frac{\me+2\mus-6\muz}{2\mm\muz}\frac{\epspzpsscm}{\rs}\rcur.  \end{eqnarray}

\end{document}